# INTEGRATED PLASMO-PHOTONIC SENSOR WITH VOLTAGE CONTROLED DETECTION

**Jacek Gosciniak[1,*] and Ryszard Piramidowicz[1]**

[1]Institute of Microelectronics and Optoelectronics, Warsaw University of Technology, Koszykowa 75, 00-662 Warsaw, Poland
Corresponding authors: * jacek.gosciniak@pw.edu.pl

## ABSTRACT

In this paper, we propose a waveguide-integrated interferometric sensor in which interference occurs between two plasmonic modes propagating in a single plasmonic waveguide. For the purpose of sensing, the vertical plasmonic slot waveguide was rearranged by increasing the distance between the metal electrodes. Consequently, the plasmonic modes associated with each metal electrode, which typically form a hybrid plasmonic slot mode, have been separated, enabling them to propagate independently on opposing edges of metal electrodes. This allows for the implementation of a Mach-Zehnder interferometer, in which light is coupled into and out of the structure from a photonic waveguide through a conventional tapered structure. Notably, the metal electrodes that support the plasmonic modes can also function as electrical contacts. By applying a DC voltage between them, it is possible to efficiently separate ions that drift to one of the metal electrodes. Consequently, one of the arms of the Mach-Zehnder interferometer experiences higher losses and a phase accumulation, providing an imbalance of the Mach-Zehnder interferometer and a drop in transmission. Here, any change in a transmission refers only to the amount of ions in a liquid as the output signal from the interferometer is normalized to a liquid by the reference arm which is in direct contact with the examined liquid solution. The total amount of ions in the examined liquid remains constant, however, what changes is their distribution in the gap as the ions drift toward one of the metal electrodes when a voltage is applied. Thus, any change in the concentration of ions in the liquid can be monitored by transmission measurements from the interferometer. The proposed configuration is highly sensitive to variations in transmission between the two arms of the interferometer, enabling a record sensitivity of over 12460 nm/RIU, even at the telecom wavelength of 1550 nm. A further enhancement in sensitivity is expected in the mid-infrared wavelengths, which correspond to the maximum absorption peaks of most chemical and biological compounds.

## Introduction

In recent years, integrated photonics sensors have attracted a lot of attention due to a high-sensitivity, label-free and real-time operation, and the potential of mass production at a low cost [**1-4**]. In terms of the operation conditions, an on-chip spectroscopy can be classified in three main categories: (a) absorption spectroscopy, (b) refractometry spectroscopy and (c) Raman spectroscopy. In the case of an absorption spectroscopy, sensor measures the change in optical power transmitted through a waveguide under an interaction of the evanescent field with the chemical or biological compounds of interest placed close to the waveguide [**2, 3, 5**]. In comparison, for refractory sensing, the sensors can be implemented either in interferometric arrangements like Mach-Zehnder configuration or in resonant structures like ring resonators (RRs). For a refractory sensing with the Mach-Zehnder Interferometer (MZI), sensor measures a change in optical power at the termination of the MZI as a result of a change in the mode effective index in one of the MZI arms due to a presence of chemical or biological compounds [**1**] or, as in a case of ring resonator (RR) based sensors, a change in the through-port power due to a presence of molecules in the cladding that change the refractive index of the surrounding, and in consequence, the ring waveguide core [**6**]. While the resonant structures like RRs can be ultra-compact, their sensitivity is limited to around 500-700 nm/RIU [**7**]. In comparison, the MZIs are more tolerant of any fabrication imperfections and more versatile while simultaneously providing much higher sensitivity values [**8-11**]. Finaly, for a Raman spectroscopy, light at an initial wavelength propagating in the waveguide interacts with a surrounding molecule and is scattered at a new wavelength. The excitation light and the spontaneous Raman emission signal from molecules co-propagate in the same waveguide while the Raman signal serves as the signal of interest that defines chemical or biological molecules [**4**].

In this paper we focus on a refractory sensing mechanism based on the MZI as we believe it is the most promising technique for the realization of sensitive and compact sensors that are able to detect a broad spectrum of chemical and biological molecules.

To ensure a full operation a sensor requires a source [**12-14**], a transducer, and a detector [**15-19**]. The optical transducer is the clue component for an integrated photonic sensor that detects some environment changes under certain, established conditions, however, it requires a light source and a detector to provide fully independent measurements for a specific wavelength range. The MIR wavelength region is under a special interest in this type of measurement as many chemical and biological molecules exhibit intense and unique absorption features in this wavelength band [**1, 2**]. However, the realization of MIR sources and their integration with an available material platform is very challenging while MIR photodetectors show poorer performances compared to their NIR counterparts due to the lower photon energy they provide. Thus, the technology behind MIR sensors is particularly challenging and requires far more research in coming years.

## Photonic vs. plasmonic transducer

In this paper we focus on a transducer as a key component of the sensor. The mechanism of operation of the waveguide-integrated photonic sensors operating based on the absorption and refractory spectroscopy typically relies on the interaction of the evanescent field of the guided optical mode with the analyte ions/molecules placed on top of the transducing photonic waveguide and detection is bound by propagation loss of the waveguide. However, the performance of such sensors is limited by a weak evanescent field from a photonic waveguide. In consequence, a long sensing arm is needed that often exceeds several millimeters or even centimeters what influence its compactness and restricts its applications. One of the solutions relies on replacing photonics waveguide by their plasmonic counterpart. As a result, a significant electric field enhancement can be achieved what translates on a stronger interaction of the electromagnetic field with ions/molecules placed in a sensing medium and, consequently, shorter interaction length. Apart from a higher electric field, the plasmonic counterparts offer higher sensitivity values as the analyte with the ions/molecules can be placed directly in contact with the metal electrode, i.e., in the maximum electric field of the operating mode, thus enhancing the interaction of the electromagnetic radiation with the ions/molecules placed in the analyte. Simultaneously, the inevitable high propagation losses of the plasmonic waveguides can be counterbalanced by co-integration of plasmonic structures that serve as transducer with low-loss photonic waveguides that deliver a light to the plasmonic transducer and then collect it from a sensor. Through such an integration the miniaturized photonic integrated circuit (PIC) for a sensing application can be delivered.

## Absorption vs. refractory sensing

The absorption sensing mechanism relies on the comparison of a difference in transmission of light through the photonic or plasmonic waveguide under the absence and presence of the sensing medium. However, such a device requires sequential reference and sensing measurements due to constant drift or fluctuations in the light source power. To reduce the time-dependent measurements errors the effective referencing strategy can be considered where the ratio of the transmission through two arms, reference and sensing, is examined, i.e., the light can be split between the reference and sensing arms while a pair of detectors can be used to continuously measure the signals from both arms [**3**]. Accordingly, the transmission through the sensing arm is normalized to the reference arm, thus, any losses not related to the sensing medium are removed and only the absorption spectrum from the sensing medium can be obtained.

To reduce complexity and increase the efficiency of a sensor, it is feasible to combine the two arms at the terminal side, thereby creating the Mach-Zehnder Interferometer (MZI) with a photodetector placed at the termination of the MZI. Consequently, the refractory sensing with the MZI can be attained by employing a photodetector to measure a change in transmitted power resulting from a change in the complex mode effective index within the sensing arm due to the presence of a sensing medium and ions/molecules in it. In the presence of a sensing medium, the mode effective index in the sensing arm undergoes a change, thereby inducing a phase shift between two propagating modes in both the reference and sensing arms. This phase change is then translated into a spectral shift of the MZI resonance, which consequently results in a change in the transmitted light. A difference in a transmitted power under an absence and presence of the sensing medium and ions/molecules located in it defines the extinction ratio (ER) of the MZI. To maximize the spectral ER and thus the sensitivity of a sensor, it is essential to equalize the power output from the two arms of the MZI. This can be accomplished either through design by making the sensing arm longer or shorter compared to the reference arm thus introducing additional losses or by placing a variable optical attenuator (VOA) in the reference arm.

**Proposed concept**

The proposed concept aims to address the various issues above simultaneously by proposing an ultra-sensitive sensor device that can be integrated into micrometer scale chip-based configuration by using simple and low-cost fabrication methods.

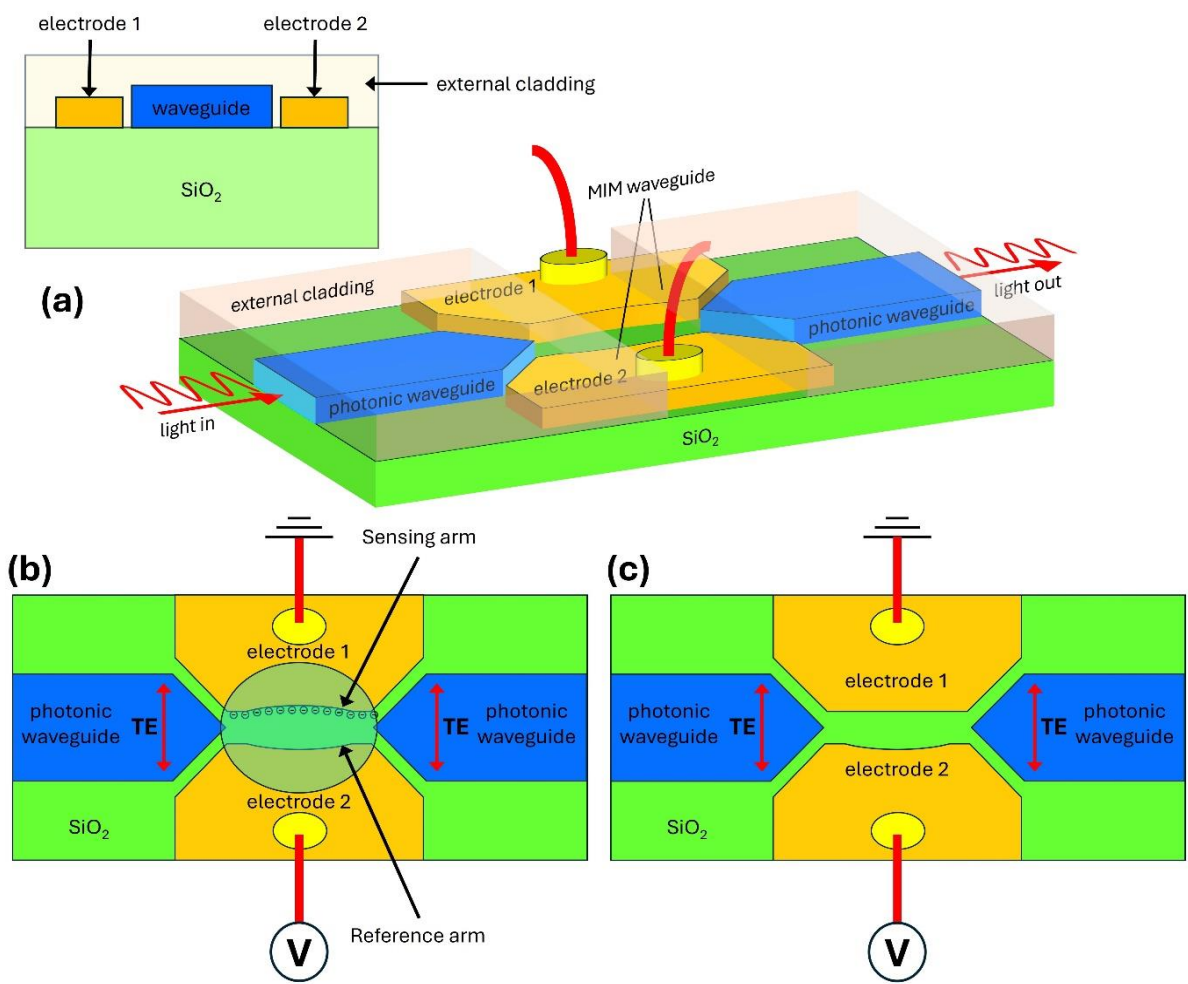

**Figure 1**. Perspective view (a) and top view (b, c) respectively of the hybrid plasmo-photonic slot waveguide arranged in the Mach-Zehnder Interferometer schema used for the plasmonic section of the sensor and the biased voltage applied between electrodes. The plasmonic slot waveguide were arranged with both curved electrodes (a, b) and one electrode straight and second curved (c) in the center of the slot.

The paper contains **two** novel concepts of on-chip sensors that allow to achieve high sensitivity at extremely short sensing length. The **first** novelty relies on the utilizing the well-known plasmonic slot waveguide (here metal-insulator-metal (MIM) waveguide) [**20, 23-26**] as the Mach-Zehnder Interferometer (MZI) where two opposite sides of metal electrodes, which constitute of a slot, are used as the MZI arms. In such an arrangement, the light is provided to the MZI "splitter" by a photonics mode where the splitter consists of a tapered photonic waveguide coupled to the plasmonic slot waveguide (**Fig. 1**). The second "splitter" that collects a light from the MZI "arms" consists of similar tapered photonic waveguide connected with a plasmonic slot waveguide.

As with other MZI-based sensors, the sensing mechanism involves a detection of refractive index change in one of the MZI "arms" that is in contact with a sensing medium. This refractive index change induces a phase change of the sensitive plasmonic mode that propagates in this arm. The MZI translates this phase change into a wavelength shift at the interferometer output, i.e., in the photonic waveguide. As both MZI "arms" in the proposed sensor arrangement are in a direct contact with a sensing medium and ions/molecules, some separation mechanism is needed to separate ions present in the medium to introduce a high gradient distribution of ions in the sample. In consequence, the second novelty is proposed in this paper.

The **second** novelty relies on an active separation technique that allows effective separation of any chemical or biological substances and/or molecules, ions in a sensing medium in response to an applied electric field between metal electrodes that, simultaneously, constitute of a plasmonic slot (MIM) waveguide and thus it allows for a stronger interaction of the plasmonic mode with ions present in the sensing medium (here water) (**Fig. 1**). In the absence of electric field in the sensing medium (here water), the positively and negatively charged ions are evenly distributed in the liquid solution. In consequence, the ions concentration close to a one of the metal surfaces i.e., in the maximum of the propagating plasmonic mode, is very low thus, the interaction of the plasmonic mode with the ions is reasonably weak and, consequently, the response of the system to the presence of the ions in a liquid is weak. However, under the applied DC voltage the negative charged ions (anions) move towards a positive electrode (anode) while the positive charged ions (cations) move towards a negative electrode (cathode). Therefore, ions are shifted to the maximum electric field of the propagating plasmonic mode, thus, the interaction of the mode with the ions is highly enhanced what translates on the stronger response of the sensor.

## Operation principles of the MZI-based sensors

For a nonresonant structures such as, for example, a straight or spiral waveguides, the sensing mechanism relies on the interaction of the sensing medium ions/molecules with the evanescent field of the guided optical mode and the sensitivity of such sensors is related to a propagation loss of the waveguide. For a $L$-long sensing waveguide that is characterized by initial losses of $a$ related to the waveguide geometry (in the absence of the ions/molecules in the sensing medium), the change of transmitted light intensity $\Delta I$ due to the presence of an ions/molecules in the sensing medium is given by

$$\Delta I = I_0 exp(-\alpha L)[1 - exp(-\Gamma \alpha' L) \approx I_0 exp(-\alpha L)] \cdot \Gamma \alpha' L \qquad (1)$$

where $\alpha'$ denotes the absorption coefficient of the sensing medium and $\Gamma$ is the modal confinement factor in the sensing medium. Thus, the sensitivity can be expressed by the change in fractional optical intensity $\Delta I/I_0$ induced by a minor change in the number of analyte ions/molecules which consequently leads to a change in the cladding absorption coefficient. Furthermore, the sensing waveguide should be characterized by very low propagation losses within considered spectral band [**2**].

On the contrary, the spectral response of a proposed MZI, as well as other MZI-based interferometric devices, varies with wavelength and depends on the beam intensity ratio between both propagating beams along the MZI arms and the phase difference between them, thus the output intensity is given by:

$$I_{out} = (A_1 + A_2)^2 = |A_1|^2 + |A_2|^2 + 2|A_1||A_2|cos(\Delta\phi) \qquad (2)$$

where $I_i=|A_i|^2 exp(i\phi_i)$ and $\phi_i=2\pi n_i L/\lambda$ (for $i$=1, 2, i.e., for both arms of the MZI). Here, $A_i$ is the mode amplitude in one of the MZI arms, $n_i$ is the effective refractive index of the corresponding amplitude, $L$ is the length of the interferometer arm and $\lambda$ is the source wavelength. In consequence we receive:

$$I_{out} = (|A_1|^2 + |A_2|^2)\left(1 + \frac{2|A_1| \cdot |A_2|}{|A_1|^2 + |A_2|^2}cos(\Delta\phi)\right) \qquad (3)$$

where $\Delta\phi=2\pi\Delta n\cdot L/\lambda$ is the phase difference between both MZI arms. Replacing beam amplitudes by the intensities we receive:

$$I_{out} = (I_1 + I_2)\left(1 + \frac{2\sqrt{I_1 I_2}}{I_1 + I_2}cos(\Delta\phi)\right) \qquad (4)$$

where factor

$$V = \frac{2\sqrt{I_1 I_2}}{I_1 + I_2} \qquad (5)$$

also known as a visibility, defines the beam intensity ratio, $I_1$ and $I_2$, between both MZI arms while $\Delta\phi$ corresponds to the phase difference between them that, for equal length of the interferometer arms, is proportional to the difference in the mode effective indices between two propagating modes, $\Delta n$.

For an equal beam intensity ratio between both MZI arms, i.e., $I_1=I_2=1/2 I_{inp}$ where $I_{inp}$ is the input intensity at the entrance of the MZI, $I_{inp}=I_1+I_2$, the visibility $V$=1 and the absolute value of a transmission from the MZI, that can be simplified using the trigonometric function of the double-angle formula $cos(2a)=2 \cdot cos^2(a)-1$, is proportional to the phase difference between both MZI arms according to the equation:

$$\frac{I_{out}}{I_{inp}} = cos^2\left(\frac{\Delta\phi}{2}\right) \quad (6)$$

In consequence, the MZI-based sensor exhibits at the output the transmission dips and peaks that corresponds to destructive and constructive interference of light propagating along the sensing and reference arms.

For an optical sensor operating based on the interferometric structures the sensitivity is defined by the amount of wavelength shift caused by the effective index change of the propagating mode along the sensing waveguide due to a change in the ambient refractive index. For a proposed MZI-based sensor the wavelength shift is caused by the change of the effective index of the propagating plasmonic mode due to exposure it to the sensing medium located in a proximity of the metallic electrode supporting the propagating mode.

**Homogeneous sensing mechanism vs surface sensing mechanism**

The sensitivity of the sensors is determined by the ratio of the resonance shift of wavelength to the refractive index change and can be calculated using the following formula:

$$S = \frac{d\lambda}{dn_{liq}} = \frac{d\lambda}{dn_{eff}} \frac{dn_{eff}}{dn_{liq}} \quad (7)$$

where $\lambda$ is the wavelength of the optical signal, $n_{liq}$ is the refractive index of the considered liquid and $n_{eff}$ is the mode effective index in the plasmonic waveguide [**21**]. Here, the first term $d\lambda/dn_{eff}$ describes the resonant wavelength shift produced by the MZI sensor as a result of the effective refractive index change of sensing waveguide which is depended on the architecture of the sensor while the second term $dn_{eff}/dn_{liq}$ describes the sensitivity of the sensing waveguide, which is proportional to the optical confinement factor $\Gamma$ of covering material (liquid) under sensing. The optical confinement factor $\Gamma$ is defined as the power confined in particular area divided by the total power produced by a waveguide. However, this is only true for sensing of chemical or physical quantities which are homogeneously distributed in the evanescent field of the waveguide (**homogeneous sensing mechanism**). In a case of sensing of ultrathin chemical, physical or biological layers of thickness much lower than a wavelength of light that are immobilized on the surface of a waveguide or in very close proximity of it, the sensitivity is defined as the rate of change of the modal effective index of the propagating mode versus the dielectric load term that is defined by a thickness of such a layer and the dielectric function difference between a thin layer and the cover material, here liquid (**surface sensing mechanism**) [**22**]. The sensitivity of the surface sensing mechanism can be enhanced by increasing the electric field enhancement in the proximity of the waveguide, i.e., in the area where a thin layer is deposited, which can be realized through plasmonic waveguide arrangements. Additionally, the sensitivity is enhanced in symmetric structures where the refractive index of the material beneath a waveguide is equivalent to the refractive index of the material above a waveguide. It is noteworthy that the surface sensing mechanism exhibits exceptional sensitivity to minor alterations in the thickness of the thin layer or its refractive index, resulting in substantial changes in the transmitted light through a waveguide [**22**].

In light of the aforementioned arguments, we have implemented the **surface sensing mechanism** in the proposed MZI-based slot waveguide.

## Proposed on-chip sensor arrangement

The schematic of the proposed waveguide-integrated sensor is based on the **surface sensing mechanism** and is shown in **Fig. 1**, where the light couples to the plasmonic section from the in-plane tapered photonic waveguide [**19, 23**]. However, it is important to note that other coupling schemes can be considered as well. For example, evanescent coupling from a photonic waveguide buried in the substrate below or directly from an optical fiber into a plasmonic device through a grating coupler can make a proposed plasmonic sensor even more compact and integrable on a variety of substrates. In the case of the in-plane tapered plasmonic waveguide, Koch [**24**] has shown that excess losses as low as 0.5 dB can be achieved. For the evanescent coupling, Salamin [**25**] has demonstrated that a perfectly fabricated device can exhibit efficiency levels of up to 90%.

The proposed sensor that is based on the plasmonic slot (MIM) waveguide offers a lot of benefits as the metal electrodes that support the plasmonic modes can serve simultaneously as high-speed electrodes what enables an optimal overlap between the electrical and optical fields (**Fig. 2**) [**24**]. The electric field enhanced in the slot enables a significant reduction of the sensor by facilitating the transfer of ions to the side of one of the metal electrodes. This, in turn, reduces both the contact resistance and the capacity of the device, as well as enhances the interaction of light with ions during measurements. Consequently, the RC time constant decreases, thereby allowing for an increase in the electrical bandwidth.

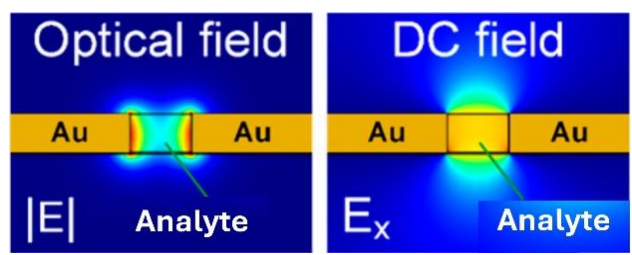

**Figure 2**. The cross-section of the simulated optical and direct current (DC) electrical fields of the plasmonic slot waveguide with a sensing medium core and Au electrodes.

The active plasmonic slot waveguide (MIM waveguide) can consist of two symmetric or asymmetric metallic contacts with sensing medium serving as the absorbing core. The coupled photons are converted into Surface Plasmon Polaritons (SPPs) that propagate along the MIM slot as two separated SPPs. For a small slot width, the SPPs that propagate on the opposite metal contact coupling together giving rise to the plasmonic gap mode.

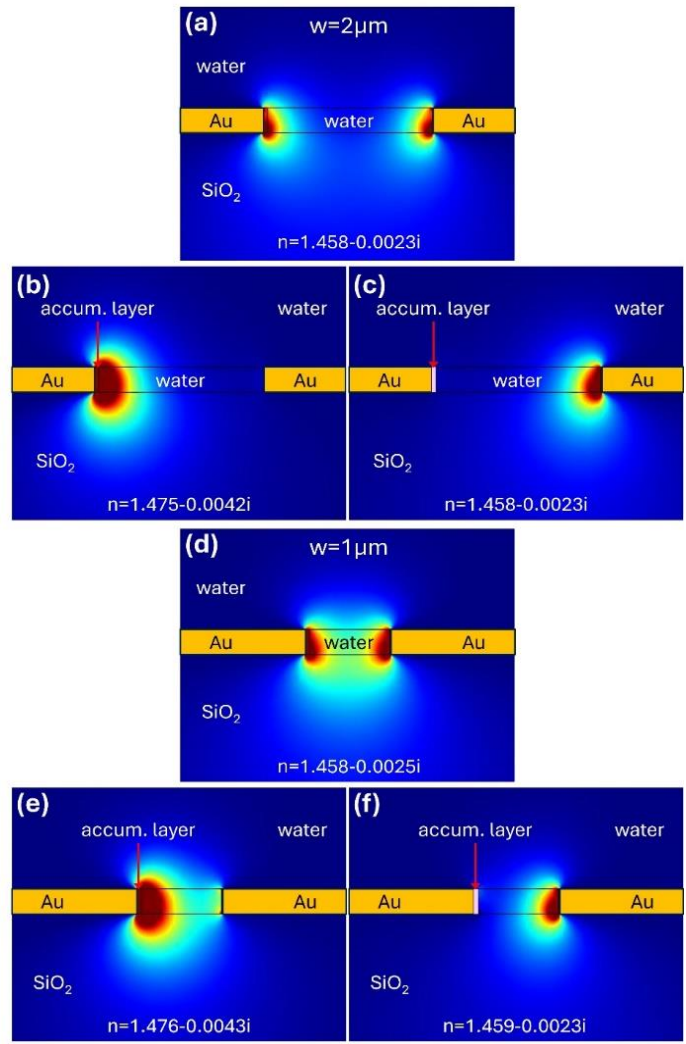

**Figure 3**. The cross-section of the simulated in-plane optical electric field component of the propagating TE mode for a plasmonic slot waveguide with a thin layer of ions/molecules accumulated in one metal-water interface and Au electrodes. Simulations were performed for a gap width of *w*=2 µm (a, b, c) and *w*=1 µm (d, e, f) and without (a, d) and with ions/molecules (b, c, e, f). The thickness of Au electrodes was kept constant at *h*=300 nm.

However, when the distance between opposite metal contacts/electrodes is efficiently high, the mode effective index approaches asymptotically the effective index of the SPP of a single metal-dielectric interface. That is to say, the gap plasmonic mode separates into two SPP modes that propagate on the opposite edges of the metal contacts (see **Fig. 3**). In a termination section of the metal structure, both SPP modes couple back through a taper to the photonic waveguide. Given that the photonic waveguide is designed to support only a fundamental TE mode, the coupled SPPs form two separated TE propagating modes in the plasmonic slot waveguide. As previously mentioned, the proposed structure functions as a Mach-Zehnder interferometer, with the MZI arms being created by metallic contacts. In the absence of an applied DC voltage, the distribution of ions within the gap remains uniform, leading to equivalent attenuation and phase shifts experienced by both SPPs propagating along opposite metal contacts. Consequently, at the structure's termination, the SPPs couple back to the photonic waveguide with the same phase and intensity. As a result, the maximum power output from the photonic waveguide can be detected. However, when a biased DC voltage is applied between the two metal contacts, a uniform electric field is generated in the gap (**Fig. 2**). The electric field generated in the gap separates positive and negative ions (anions and cations) present in the sensing medium (here water), directing them towards opposite metal electrodes. Consequently, positive ions migrate towards a metal electrode with a negative DC voltage applied to it, while negative ions migrate towards a metal electrode with a positive DC voltage. This results in the ions settling on one of the metal electrodes or in close proximity to it, while the other electrode remains ions-free (**Fig. 4**). This results in a state of imbalance in the MZI formed by two SPPs modes propagating on the opposite side/edges of the metal electrodes. The ions that are deposited on the side of the electrode give rise to higher absorption and, simultaneously, induce a refractive index change of the propagating SPP mode associated with this electrode, which in turn induces a phase change between both MZI arms. Consequently, when both SPPs couple back to the photonic waveguide, the transmission decreases. In the extreme case, for a $\pi$ phase shift between both SPPs modes, the transmission reaches a minimum. In conclusion, the change of the mode effective index of one of the plasmonic modes that constitute the MZI arms results in a shift of the spectra resonances of the MZI, as it is extremely sensitive to the medium residing in the gap and can serve for monitoring the amount of different ions present in a liquid.

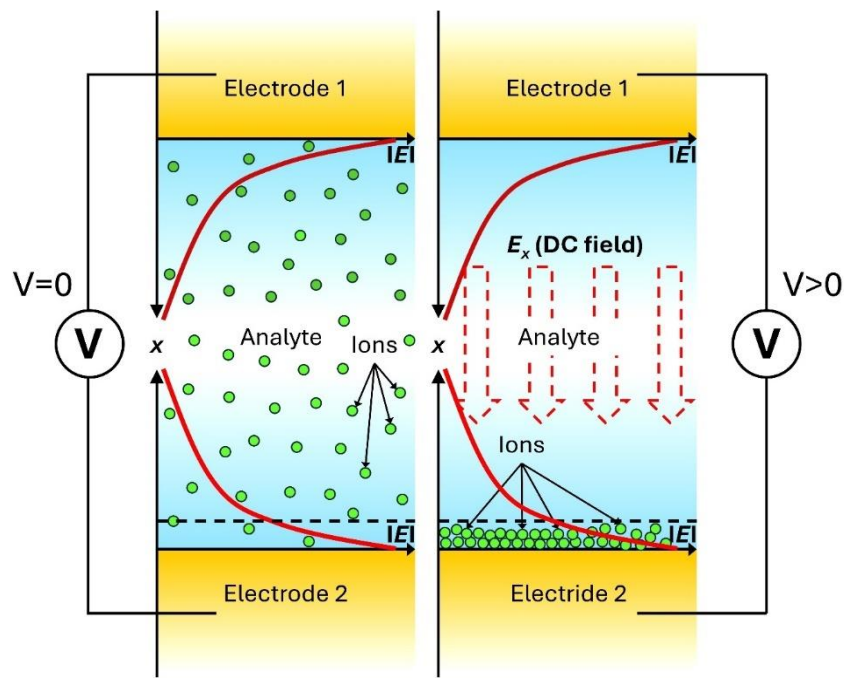

**Figure 4**. The top view of the plasmonic slot waveguide with the optical and direct current (DC) electric field distributions in the sensing medium, here water, under zero applied voltage (a) and an applied voltage (b) with the characteristic ions' distribution.

The operating principles of the proposed sensor arrangement were initially simulated at a telecom wavelength of 1550 nm, however, the real device should operate at wavelengths that correspond to the maximum absorption peaks for a given chemical compound as shown in **Fig. 5**. Since most of the chemical compounds show the characteristic absorption peaks in the MIR, the proposed sensor should be designed for these specific wavelengths.

The simulations were performed for the Au gold electrodes with a separation between them of 2 µm (**Fig. 3a, b, c**) and 1 µm (**Fig. 3d, e, f**). The thickness of the Au electrodes was kept constant at 300 nm to match the thickness of the SiN photonic waveguide, which was assumed to couple a light to a considered sensor. In addition, the simulations were performed for water as a test material with some ions in it with a refractive index corresponding to $n$=1.8. The water with a mixture of ions was placed directly in the gap and on top of the MIM plasmonic structure.

At zero applied voltage, the SPP modes at both edges of the metal electrodes are the same, so they couple together to provide one MIM plasmonic mode with a complex mode effective index $n_{eff}$=1.458+0.0025·$i$ for a metal separation of 2 µm and $n_{eff}$=1.458+0.0028·$i$ for a metal separation of 1 µm. Those values correspond to the attenuation losses of 0.09 dB/µm and 0.10 dB/µm for metal separations of 2 µm and 1 µm, respectively. The higher losses for a smaller separation between metal electrodes are related to the higher absorption by metal electrodes, since more electric field penetrates a metal resulting in higher attenuation. In such a case, the beam intensity ratio in both MZI arms is the same and both SPP modes encounter the same phase shift and attenuation, so there is not a phase shift between both MZI arms. Consequently, the symmetric structure without an applied DC volage behaves as a nonresonant structure. As it has been previously shown experimentally, the transmission spectrum of such a MIM waveguide shows a flat response over a broad spectral range and depends only on the length of the plasmonic MIM waveguide [**26, 29**]. However, when the DC voltage is applied between metal electrodes the ions start to move towards one of the metal electrodes, i.e., the negative ions drift towards the metal electrode with a positive DC voltage while the positive ions, if present in the liquid at the same time, drift towards the metal electrode with a negative DC voltage applied to it. As a result, ions settle on one of the metal electrodes while the other remains free of ions. This introduces an asymmetry between the two arms of the MZI, which depends on the ion concentration on the electrode and the optical properties of the ionic compounds for a given wavelength of interest. Assuming the accumulation layer thickness of 50 nm that is created when ions move towards one of the metal electrodes and the refractive index of ions under an interest of $n$=1.8, the mode effective index for the SPP mode associated with this interface was calculated to be $n_{eff}$=1.475+0.0042·$i$ while for a second interface without ions it was calculated to be $n_{eff}$=1.458+0.0023·$i$ which translates into propagation losses of 0.15 dB/µm and 0.08 dB/µm, respectively (**Fig. 3**). Based on a provided data, the $L$=45.5 µm long MIM waveguide is required to provide a π phase shift between both arms of the interferometer with a *V* factor (visibility) calculated for this length of MIM waveguide at $V \approx 0.94$, which translates to the extinction ratio ($ER=I_{max}/I_{min}$) exceeding an impressive value of 15 dB. Furthermore, the calculation has shown that for a 100 μm long MZI waveguide arranged under the same conditions, a phase shift of more than 2π between both MZI arms is possible with the visibility exceeding $V=0.75$ which gives an extinction ratio $ER=8.5$ dB.

As the concentration of ions under a test in a liquid increase, more ions move to one of the metal electrodes under an applied DC voltage. Consequently, the concentration of ions on this metal electrode increases, thereby enhancing the interaction of ions with the electromagnetic field supported by one of the propagating SPP modes.

## Influence of the absorption spectrum of the sensing medium on the choice of wavelength spectrum

Each chemical or biological compounds show characteristic absorption spectrum with the maximum absorption that originates from stretching vibrations and bending modes of the molecules that constitute for a given compound [**27**]. The maximum absorption corresponds to the highest losses, thus, the material is characterized by the highest imaginary part of the permittivity (**Fig. 5**). In consequence, a simple straight waveguide that operates based on the change in transmission of light through the sensing medium because of the high absorption of a sensing medium can be considered [**5**]. However, at the same time, the maximum absorption corresponds to the highest change of the real part of permittivity what enables the construction of sensors that operate based on the refractive index change what is directly corelated with the change of the phase of the propagating mode through the sensing medium. For this second scenario, the resonant devices such as Mach-Zehnder Interferometers (MZIs) or ring resonators (RRs) can be considered [**1, 3, 9**]. To date, numerous studies have shown that MZI is more tolerant of minor inaccuracies during the production process while it can offer higher sensitivity values compared to RR-based sensors [**1, 3**]. Up to now, several different MZI arrangements were proposed to maximize sensitivity, minimize the footprint and the overall complexity [**1, 3, 21, 28**]. While most of the proposed MZI sensors operate based on the two spatially separated arms with one serving as a sensing arm and the second as a reference arm [**1, 3, 28**], more sophisticated designs were proposed as well. One of them, known as bi-modal interferometers, utilizes a single-path interferometer that exploit modes with different properties propagating in the same physical channel, i.e., on top and bottom layer of the metal stripe, which interfere together at the termination of the metal stripe [**10, 11**]. In this arrangement the bottom metallic surface plays the role of the reference arm of the interferometric structure while the top metallic surface is exposed to the liquid under test and, thus, serves as the sensing arm. However, such sensor design requires very precise alignment of the metal stripe in relation to the input and output waveguide and even a small deviation from an optimal alignment can cause a huge change in the coupling ratio between both modes. Furthermore, for each wavelength under an interest, the bi-modal structure should be redesigned as it will require a different thickness or even material of the bottom layer to match the refractive index of a sensing medium to a desired wavelength. In consequence, it will require realignment of the metal stripe relative to the input and output waveguides. As a result, this sensor design is extremally challenging for manufacturing and, in general, impractical for applications where a broad spectrum of wavelengths is considered.

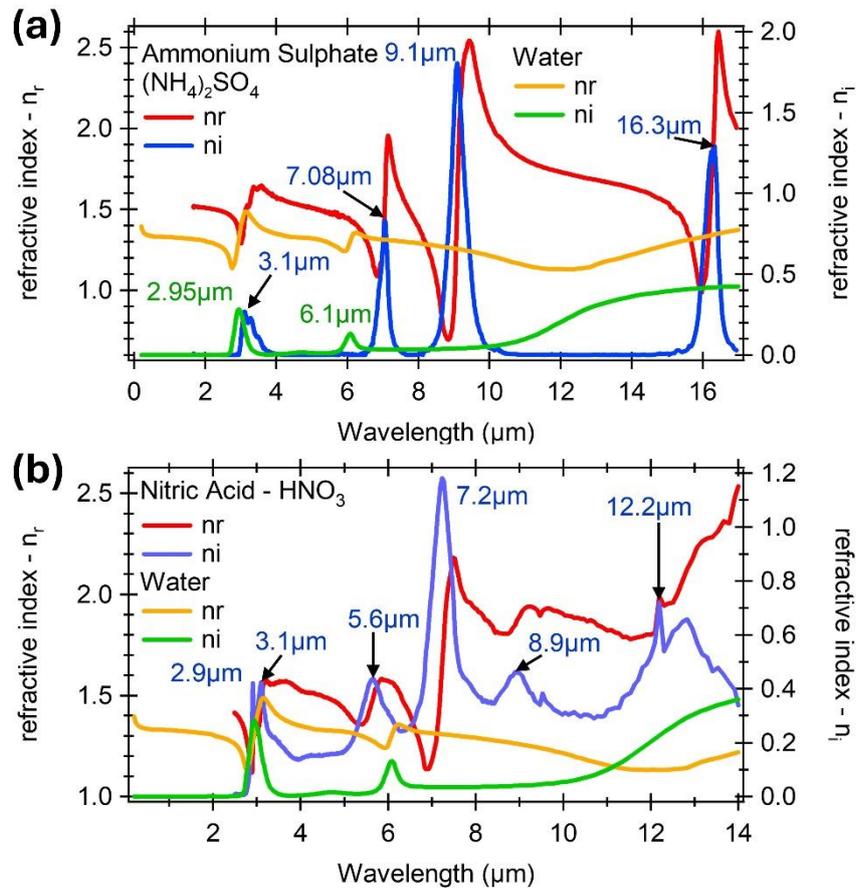

**Figure 5**. Example of a complex refractive indices of water and some chemical compounds such as (a) $(NH_4)_2SO_4$ and (b) $HNO_3$ with some characteristic absorption peaks corresponding to the maximum of the imaginary part of complex refractive indices [**30**].

The proposed sensor is capable to detect a broad range of common ions present in water and aqueous solutions including nitrites ($NO_2^-$), nitrates ($NO_3^-$), phosphates ($PO_4^{3-}$), and ammonium ions ($NH_4^+$), among others. This makes it an ideal candidate for monitoring water pollutants. To fully leverage the proposed design, it is essential to operate within the spectral range where ions exhibit the highest absorption peaks, thereby maximizing sensitivity. The mid-IR band is of great significance, encompassing the primary absorption bands of the majority of chemical and biological molecules, as well as the fingerprint region (7-20 μm). This makes it a crucial component in spectroscopic sensing. The mid-IR absorption of most chemical compounds is at least 100 times stronger than their absorption bands in the near-IR [**1, 3, 30**] despite the lower energy they carry in comparison to their visible or near-IR counterparts. This makes them more susceptible to thermal fluctuations.

The proposed sensor design is very sensitive to the presence of ions in a liquid solution thus it can operate even in spectral range that is far away from a desired wavelength range due to the surface sensing mechanism implemented in a proposed structure. The 2D eigenmode simulation performed at wavelength of 1550 nm revealed that for $L$=150 μm long sensor and an Au metal electrode separation of 2 μm deposited on SiO$_2$ substrate ($n_{SiO2}$=1.45) and for a water filling a gap between metal electrodes, that is described by a complex refractive index $n_{water}$=1.318+9.8625·10$^{-5}i$ with NH$_4^+$ ions ($n_{NH4}$=1.517) bounded to one of the metal electrodes (**Fig. 5**) that thickness change with the concentration of ions in a water, the surface SPP sensitivity of this mode was found equal to 0.00008025 RIU/nm (**Fig. 6**) according to the equation

$$S_s = \frac{\delta n_{eff}}{\delta t} \qquad (8)$$

Here, $n_{eff}$ represents the effective index of the SPP mode with bounded ions and $t$ is the thickness of the ions layer. In consequence, a sensor surface sensitivity was calculated at 12460 nm/RIU what is at least **3 times** higher than sensitivities previously reported for a bi-modal sensor configuration [**10, 11**] and more than **one order** of magnitude (10 times) higher than for other proposed sensors [**11**].

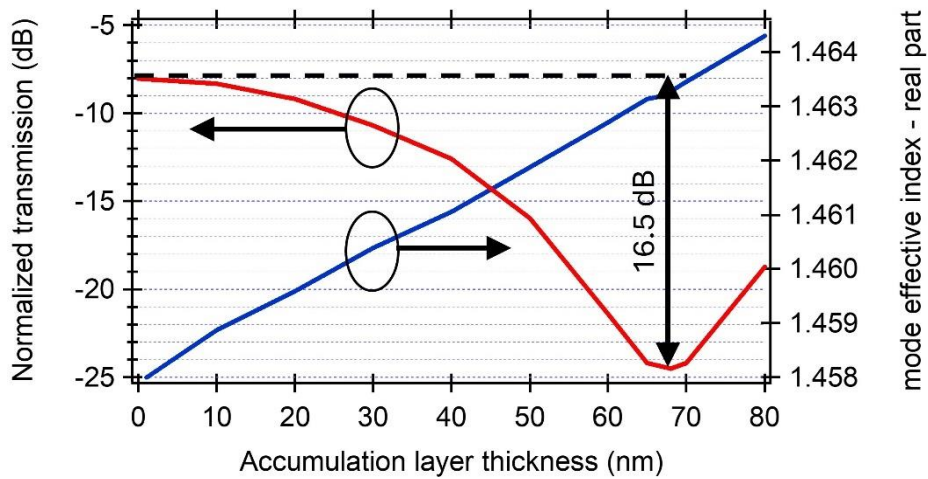

**Figure 6**. Normalized transmission and mode effective index (real part) of a 150 μm long sensing arm of the MZI as a function of the accumulation layer thickness, i.e., the NH$_4$ ions thickness, for the Au electrode separation of 2 μm.

Concurrently, the extinction ratio (ER) between two distinct conditions, namely the absence of ions in water and a 68 nm-thick layer of ions bound to one of the metal electrodes, was calculated to be 16.5 dB. For this thickness, a difference in the phase between both MZI arms reaches $\pi$, and a light transmitted through a sensor is minimal. It is important to note that further increases in the thickness of the ions will lead to further increases in the phase difference between both MZI arms, which will result in a subsequent increase in the transmitted light from the interferometer as observed in **Fig. 6**.

The results presented herein were achieved for a low refractive index contrast between water and NH$_4$ ions calculated at $\Delta n \approx 0.2$, where the refractive indices of water and NH$_4$ ions were taken at $n_{water}$=1.318 and $n_{NH4}$=1.517, respectively, for an operation wavelength of 1550 nm. However, by approaching the absorption peaks of NH$_4$ ions, defined by the maximum value of its imaginary part of the refractive index, the refractive index contrast between water and NH$_4$ increases significantly (**Fig. 5a**). Consequently, for an operation wavelength of approximately 7.15 µm, the refractive index of NH$_4$ ions is $n_{NH4}$=1.96, while for water it is defined at $n_{water}$=1.31. This results in a refractive index contrast of $\Delta n \approx 0.65$, which is more than three times higher than that at the examined wavelength of 1550 nm. Even higher contrast of $\Delta n \approx 1.3$ is achieved for a longer wavelength of 9.43 µm, where the refractive index of NH$_4$ ions is $n_{NH4}$=2.54 whereas for water it is $n_{water}$=1.24. This finding indicates that a substantial reduction in accumulation layer thickness can be attained for longer wavelengths, leading to a $\pi$ phase shift between the two MZI arms. Consequently, this enables the subsequent attainment of a one-order-of-magnitude enhancement in the sensor surface sensitivity. Consequently, the sensor surface sensitivity can exceed $10^5$ nm/RIU, which is a remarkable achievement.

In comparison with alternative refractory sensing configurations [3, 9-11], the proposed sensor offers a distinct advantage by providing solely information regarding the quantity of ions in a liquid and avoiding any other misleading information not related to the examined ions. The objective can be accomplished as both arms of the MZI are in direct contact with the liquid under test with a random distribution of ions in it, thereby enabling the sensing arm to be perfectly normalized to the reference arm. Therefore, it is not necessary to match the refractive index of the liquid by other material placed in the reference arm, as is the case of other arrangements [3, 9-11]. In the aforementioned initial conditions, the MZI is perfectly balanced, exhibiting the same beam intensity ratio in both arms and the absence of any phase shift between both propagating modes. Applying a DC voltage between metal electrodes forces ions to drift to one of the electrodes while the second remains in contact with the surrounding liquid. The number of ions that drift to one of the electrodes is contingent on the ions' concentration in the liquid and the magnitude of the DC voltage. Consequently, the transmission through the sensing arm changes, leading to an imbalance of the MZI, which results in a decrease in transmitted power at the MZI's termination point. This decline in transmission, in turn, offers insight into the concentration of ions present in the examined liquid.

## Conclusion

The main goal of each sensor is to maximize the sensitivity and thus the extinction ratio (ER) at the output of the sensor while minimizing the overall losses. As such, the proposed sensor arrangement can provide a record high sensitivity exceeding 12460 nm/RIU even at telecom wavelength of 1550 nm due to placing a sensing material, ions, in the maximum of the propagating mode. It is achieved by applying a voltage between metal electrodes that force ions to drift towards one of the electrode's edges i.e., at the maximum of the propagating plasmonic mode. Furthermore, the above concept allows to screen the photonics part of the sensor by placing it in the substrate below, thus, the entire structure can be covered with a sensing medium, here water, without a need of screening the photonic part from liquid. In such an arrangement, the transfer of energy between a photonic waveguide and a plasmonic section takes place through the evanescent coupling between photonic and plasmonic waveguides.

## Acknowledgements

The work is supported by the National Center for Research and Development within the project HYDROSTRATEG1/000E/2022 "Development of an innovative photonic water resource monitoring system (FOSMO)". The authors acknowledge the constant support of Warsaw University of Technology, Poland for the completion of this work. Furthermore, J.G. is very thankful to Prof. D. G. Misiek for his support and very valuable suggestions.